\documentclass[intlimits,twoside,a4paper]{article}

\usepackage[cp1251]{inputenc}
\usepackage[usenames]{color}
\usepackage[normalem]{ulem}

\usepackage[eqsecnum]{cmpj3}

\issue{2022}{25}{2}{23501}
\doinumber{10.5488/CMP.25.23501}
\title[Charge distribution in an inhomogeneous solid electrolyte]%
{Charge and electric field distributions in the interelectrode region of an inhomogeneous solid electrolyte}
\author[I.~Kravtsiv, G.~Bokun, M.~Holovko, N.~Prokopchuk, D.~di~Caprio]{I.~Kravtsiv\orcid{0000-0001-7344-9528}\refaddr{label1}, G.~Bokun\refaddr{label2}, M.~Holovko\orcid{0000-0001-8114-5356}\refaddr{label1}, N.~Prokopchuk\refaddr{label2}, D.~di~Caprio\orcid{0000-0001-6239-7427}\refaddr{label3}}
\addresses{
	\addr{label1} Institute for Condensed Matter Physics of the National Academy of Sciences of Ukraine,
	1 Svientsitskii Str., 79011, Lviv, Ukraine
	\addr{label2}Belarusian  State Technological University, 13a, Sverdlov str., 220006, Minsk, Belarus
	\addr{label3} Institute of Research of Chimie Paris, CNRS-Chimie ParisTech, 11, rue Piere et Marie Curie, 75005, Paris, France
}
\date{Received November 06, 2021, in final form May 15, 2022}

\begin{document}

\maketitle

\begin{abstract}
	
A solid ionic conductor with cation conductivity in the interelectrode region is studied. Due to their large size, the anions are considered fixed and form a homogeneous neutralizing electric background. The model can be used to describe properties of ceramic conductors. For a statistical mechanical description of such systems, which are characterized by short-range Van der Waals interactions and long-range Coulomb interactions, an approach combining the collective variables method and the method of mean cell potentials is used. This formalism was applied in our previous work  [Bokun G., Kravtsiv I., Holovko M., Vikhrenko V., Di Caprio D., Condens. Matter Phys., 2019, \textbf{29}, 3351] to a homogeneous state and in the present work is extended to an inhomogeneous case induced by an external electric field. As a result, mean cell potentials become functionals of the density field and can be described by a closed system of integral equations. We investigate the solution of this problem in the lattice approximation and study charge and electric field distributions in the interelectrode region as functions of plate electrode charges. The differential electric capacitance is subsequently calculated and discussed.                	

\end{abstract}

\section{Introduction}

We would like to dedicate this paper to the memory of our good friend and colleague Vyacheslav Vikhrenko who passed away nearly a year ago. Together, we had an opportunity to work and discuss many aspects of the theory of solid electrolytes including the charge and electric field distributions in the interelectrode region. We should note that we also discussed with Vyacheslav the general ideas of the present paper at the initial stage of these studies.

Solid ceramic materials remain to be an area of intensive scientific activities. They are widely used in electrochemical sensors, rechargeable batteries, supercapacitors, memory devices, fuel cells and other technological applications~\cite{Fergus2009,Gur2016,Yao2016}. These materials are characterized by high defect crystal structure and {inhomogeneities} inducing the emergence of electrical double layers close to intergrain boundaries~\cite{Lu2013,Mahato2015,Canchaya2015,Kim2015,Liu2015}. In consequence, atomic level description of charge transfer in such systems proves to be a challenging problem. In this paper we restrict out treatment of this problem to the case of {inhomogeneities} caused by the electrical field of confining electrodes and neglect the adsorption phenomena on the surface of the solid electrolyte connected with specific interactions between ions and electrodes.

A peculiar feature of the materials under consideration is the presence of anions and cations that are drastically different in size. This fact allows us to apply the adiabatic approximation and reduce the problem to a one-component model. In the framework of this description, we consider thermal movement of cations against a homogeneous negatively charged background of fixed anions confined by the crystal lattice~\cite{Fergus2009,Gur2016,Yao2016,Lu2013,Mahato2015,Canchaya2015,Kim2015,Liu2015}. To this end, we apply a cell model in the framework of lattice description whereby we consider only the case of cation leaps between lattice sites.  

Out of the theoretical approaches fit to study the ionic systems, notably the density functional theory~\cite{Mier1991,Reszko2005,Pizio2005,Jiang2011,Pizio2012,Henderson2012,Neal2017}, field theoretical methods~\cite{DdiCaprio03,DdiCaprio98,DdiCaprio05,DdiCaprio11}, the collective variables method~\cite{Yukhnovskii80, Golovko1985, soviak}, and mean field theories, the latter remains the most popular one. The approach employed in this paper is self-consistent and takes into account pair correlations and long-range interactions as well as single particle potentials forming the mean field.

Solid ionic systems with conductivity induced by ions of the same sign~\cite{Bokun2019} are characterized by short-range Van der Waals attraction and long-range Coulomb repulsion. They can also be viewed as systems interacting via a so-called SALR (short-range
attractive and long-range repulsive) potential~\cite{Ciach10,Ciach01,Archer2007,Zhuang2016,Pini2017}. A special feature of our model is that the repulsive part of the SALR potential is of Coulomb nature. Solid electrolytes are characterized by high polarizability and high values of permittivity. As a result, the energy of short-range Van der Waals interaction is comparable to the energy of Coulomb interaction~\cite{Lu2013,Mahato2015,Canchaya2015,Kim2015,Liu2015}. To address this fact, we combine two methods that had previously been used to describe systems of neutral and charged particles, namely the methods used in~\cite{Bokun2019,Rott1975,Rott79,Bokun18a,Patsahan19,Bokun18} with the collective variables approach capable of taking into account long-range interactions~\cite{Yukhnovskii80, Golovko1985}. 

In our previous paper~\cite{Bokun2019}, the homogeneous case was considered. Such a case appears in the absence of an electric field when local charge compensation takes place. Due to an external electric field, the cations in the interelectrode region cause an inhomogeneous distribution of mobile charges and the electric field. 
To describe the solid nature of the objects under study, the partition function is expanded with respect to mean potentials~\cite{Bokun18a,Rott79}. The approach~\cite{Bokun2019} applied to homogeneous systems here is extended to inhomogeneous states emerging under an external field induced by the electrodes. In the present work the results~\cite{Patsahan19,Bokun18} are generalized by taking into account the effect of correlations. In contrast to other approaches such as the density functional theory, where a density field functional is employed in combination with correlation functions, in this work we solve integral equations for distribution functions at a given density. This approach allows us to avoid the procedure of interpolating the distribution function for two different densities by two functions for respective densities of homogeneous states. This means that for an inhomogeneous system, the distribution functions are functionals of the density field. Hence, in this paper we develop a procedure for the construction and calculation of such functionals. Furthermore, we calculate thermodynamic potentials as functionals of a given density field. Unlike the approaches aimed at a fast direct calculation of experimentally measurable quantities based on the model functionals, the method proposed herein involves solving a system of integral equations. To this end, we resort to a lattice approximation which reduces the equations under consideration to a system of algebraic equations coupled only to the nearest neighbors of a given site.

The paper is arranged as follows. The generalization of the cell theory for the description of an inhomogeneous system is presented in section 2. In section 3 the reference distribution for the inhomogeneous case is considered. In section 4 the solution of the system of equations for the potentials and the lower order correlation functions in the lattice aproximation are discussed.
In section 6 the charge and electric field distributions in the region between two plate electrodes are considered. In section 7 the results obtained are applied for the calculation of  the electric capacitance. Finally, we conclude the paper by a few remarks in section 8.

\section{Cell theory for the description of inhomogeneous systems}        

We partition a system of volume $V$ containing $N$ particles into $M$ cells ($M\geqslant N$) of volumes $\omega=V/M$. Each cell can be vacant or occupied by one or more particles. The state of some cell $i$ is characterized by the occupation numbers $\alpha_i=0_i,1_i,2_i,...$ subject to the condition
\begin{align}
	\sum\limits_{i=1}^M\alpha_i=N,\qquad c=\frac{N}{M},
\end{align}
where $c$ is the mean concentration of particles in the system. 

In contrast to our previous work~\cite{Bokun2019}, here we consider the case of an inhomogeneous system characterized by an inhomogeneous distribution of the density field. To model such a distribution, we take the mean occupation number to be different and extend the number of possible occupation numbers by describing the systems in terms of quasiparticles. We consider the case of an inhomogeneous medium, whereby the concentration is a function of the cell position and is denoted  by $c_i$. The distribution of particles inside a micro cell $\omega_i$ is characterized by a generalized coordinate $q_{\alpha_i}$ that determines the potential of interaction between quasiparticles corresponding to different occupation numbers of the cell. Hence, each occupation number corresponds to a certain sort of quasiparticles. Particles of the first sort with $\alpha_i=0$ corresponding to vacancies neither interact with one another nor with the particles of other sorts. Particles of the second sort with~$\alpha_i=1_i$ interact with one another and with other particles as real particles. The case of $\alpha_i=2_i$ therefore corresponds to two real particles such that interaction of a particle of the third sort ($\alpha_i=2$) with a particle of the second sort ($\alpha_j=1$) corresponds to a system consisting of two real particles with {coordinates} $q_{1_i}$ and $q_{1^{\prime}_i}$ interacting with a particle of the {Cartesian} coordinate~$q_{1_j}$. Such a description means that we consider the cell as a box with three different states: $\alpha_i=0$ --- no particles in the cell, $\alpha_i=1$ --- only one particle in the cell and $\alpha_i=2$ --- two particles in the cell. As a result, particles of the third sort have the energy of self-interaction $U(q_{2_i})$ equal to the energy of interaction $\Phi(q_{1_i},q_{1^{\prime}_i})$ of two real particles located in cell $i$ in the states $q_{1_i}$ and $q_{1^{\prime}_i}$, respectively. 

The potential energy of the system characterized by parameters $\alpha_i$ that describe the distribution of particles across the system can be written in the form
\begin{align}
	H(q_{\alpha_1},...,q_{\alpha_i},...,q_{\alpha_M})=H_M=\sum\limits_{i=1}^M U(q_{\alpha_i})+\frac{1}{2}\sum\limits_{i=1}^M\sum\limits_{j=1}^M U(q_{\alpha_i},q_{\alpha_j}),
\end{align}
where $U(q_{\alpha_i})=\Phi(q_{\alpha_i})+U^{\rm{ext}}(q_{\alpha_i})$, and
$U^{\rm{ext}}(q_{\alpha_i})$ is the external field potential.
\begin{align}
	U(q_{\alpha_i},q_{\alpha_j})=    
	\begin{cases}
		0, & \alpha_i\alpha_j=0,\\    
		\Phi(q_i,q_j), & \alpha_i\alpha_j=1,\\
		\Phi(q_i,q_j)+\Phi(q_i,q_{j^{\prime}}), & \alpha_i\alpha_j=2,\\
	 	\Phi(q_i,q_j)+\Phi(q_i,q_{j^{\prime}})+\Phi(q_{i^{\prime}},q_j)+\Phi(q_{i^{\prime}},q_{j^{\prime}}), & \alpha_i\alpha_j=4.  
	\end{cases}
\end{align}
Now, we assume that the energy consists of the short-range and the long-range parts
\begin{align}
	\Phi(q_{\alpha_i},q_{\alpha_j})=\Phi_{sh}(q_{\alpha_i},q_{\alpha_j})+\Phi_{L}(q_{\alpha_i},q_{\alpha_j}),
\end{align}
where $\Phi_{sh}$ is the Van der Waals interaction which can be described, for instance, by the Lennard-Jones potential; $\Phi_{L}$ is the long-range Coulomb potential. Thermodynamic properties of the system can be found by expanding the configuration integral on the states of the reference system, which includes the long-range component of the energy, the energy of the external field, and the field of mean potentials corresponding to short-range interaction responsible for forming the crystal state of the system.

\section{Reference distribution of the inhonomogeneous system}

The Hamiltonian of the reference system can be written in the form
\begin{align}
	\label{ham_ref}
	H_0(q_{\alpha_1},q_{\alpha_M})=H_0=\sum\limits_{i=1}^M U^{\rm{ext}}(q_{\alpha_i})+\sum\limits_{i=1}^M\sum\limits_{j(i)}^M \phi_{ji}(q_{\alpha_i})\nonumber\\
	+\frac{1}{2}\sum\limits_{i=1}^M\sum\limits_{j=1}^M \Phi_{L}(q_{\alpha_i},q_{\alpha_j})+\sum\limits_{i=1}^M U(q_{\alpha_i}),
\end{align}
where $\sum\limits_{j(i)}$ denotes summation over the cells surrounding cell $i$.

The potentials $\phi_{ji}(q_{\alpha_i})$ present in equation~(\ref{ham_ref}) have the meaning of the potential of the single-particle mean force acting by the particles distributed in cell $j$ on a quasiparticle of sort $\alpha$ fixed in cell~$i$ in the state $q_{\alpha_i}$. In contrast to the case we had investigated before, whereby the potentials $\phi_j(q_{\alpha_i})$ are functions of the mean density of particles, here $\phi_{ji}(q_{\alpha_i})$ are functionals of the density field described by the values of parameters $c_{\alpha_i}$. These quantities have the meaning of mean densities of quasiparticles of sort $\alpha$ in cell~$i$. Due to the normalization condition, we have 
\begin{align}
	\sum\limits_{\alpha_i=0}^2 c_{\alpha_i}=1.
\end{align}           
At a given density field, the distribution of quasiparticles across cell $i$ is characterized by the singlet distribution function
\begin{align}
	\label{singlet_f}
	F_1^0(q_{\alpha_i})=\frac{c_{\alpha_i} \exp\left\{-\beta\left[\sum\limits_{j(i)}^M\phi_{ji}(q_{\alpha_i})+U(q_{\alpha_i})+U^{\rm{ext}}(q_{\alpha_i})\right] \right\} }{Q_{\alpha_i}},
\end{align}         
where 
\begin{align}
	\label{conf}
	Q_{\alpha_i}=\int_{\omega_i}\exp\left\{-\beta\left[\sum\limits_{j(i)}^M\phi_{ji}(q_{\alpha_i})+U(q_{\alpha_i})+U^{\rm{ext}}(q_{\alpha_i})\right]\right\}.
\end{align}
Similar to~\cite{Bokun2019}, the pair distribution function can be presented in the form
\begin{align}
	\label{pair_f}
		F_2^0(q_{\alpha_i},q_{\alpha_j})=F_1^0(q_{\alpha_i})F_1^0(q_{\alpha_j})\exp\left[-\beta\sigma(q_{\alpha_i},q_{\alpha_j})\right]S_j^{-1}(q_{\alpha_i})S_i^{-1}(q_{\alpha_j}),
\end{align}
where $\sigma(q_{\alpha_i},q_{\alpha_j})$ is the screening potential~\cite{Bokun18}. The correction factors $S_j^{-1}(q_{\alpha_i})$ can be determined from the system of equations
\begin{align}
	\label{system1}
	S_j(q_{\alpha_i})=\sum\limits_{\alpha_j=0}^2\frac{c_{\alpha_j}}{Q_{\alpha_j}}\int_{\omega_i}\frac{1}{S_i(q_{\alpha_j})}\exp\left[-\beta\sigma(q_{\alpha_i},q_{\alpha_j})-\beta\sum\limits_{j(i)}^M\phi_{ji}(q_{\alpha_j})\right]\nonumber\\
	\times\exp\left[-\beta U(q_{\alpha_i})-\beta U^{\rm{ext}}(q_{\alpha_j})\right]\rd q_{\alpha_j}.
\end{align}
Extending the derivations performed in~\cite{Bokun2019} to the present case, we see that the mean field potentials $\phi_{ji}(q_{\alpha_j})$ forming the crystal state can be found from the system of equations   
\begin{align}
	\label{system2}
	\exp\left[-\beta\phi_{ji}(q_{\alpha_i})\right]S_j(q_{\alpha_i})=\sum\limits_{\alpha_j=0}^2\frac{c_{\alpha_j}}{Q_{\alpha_j}}\int_{\omega_j}K(q_{\alpha_i},q_{\alpha_j})S_i^{-1}(q_{\alpha_j})	\nonumber\\
	\times\exp\left[-\beta U(q_{\alpha_j})-\beta U^{\rm{ext}}(q_{\alpha_j})\right]\exp\left[-\beta\sum\limits_{k\ne i,j}\phi_{kj}(q_{\alpha_j})\right]\rd q_{\alpha_j},
\end{align}
where
\begin{align}
	\label{K}
	K(q_{\alpha_i},q_{\alpha_j})=\exp\left[-\beta\Phi_{sh}(q_{\alpha_i},q_{\alpha_j})-\beta\sigma(q_{\alpha_i},q_{\alpha_j})\right].
\end{align}
The relationships~(\ref{system1})--(\ref{K}) make up a closed system of integral equations for the potentials $\phi_{ji}(q_{\alpha_i})$ and the normalization factors~$S_j(q_{\alpha_i})$. The latter makes it possible to find the singlet and the pair distribution functions of the reference system and subsequently calculate the internal energy and the Helmholtz free energy of the system.

Following the derivations presented in~\cite{Bokun2019}, one can write the free energy of the system as
\begin{align}
	\label{hey39}
	-\beta F = \sum\limits_{i=1}^M\sum\limits_{\alpha_i=0}^2\left[c_{\alpha_i}\ln\frac{Q_{\alpha_i}}{c_{\alpha_i}}+V_{\alpha_i}^{(\beta)}\right],
\end{align}    
where
\begin{align}
	V_{\alpha_i}^{(\beta)}=\frac{1}{2}\int\limits_{0}^{\beta}\rd \beta\sum\limits_{j(i)}^M\sum\limits_{\alpha_j=0}^2\int_{\omega_i}\int_{\omega_j}\Phi(q_{\alpha_i},q_{\alpha_j})F_2^0(q_{\alpha_i},q_{\alpha_j})\rd q_{\alpha_i}\rd q_{\alpha_j}.
\end{align}

\section{The solution of the system of equations for the potentials and the lower order correlation functions in the lattice approximation}

In the framework of the lattice approximation, we take into account only the locations of the particles corresponding to the minimum of the potential energy of the system. In the crystal state this corresponds to the case when particles are located in the cell sites at $\alpha=0,1$ and between the sites at $\alpha=2$ ($q_{\alpha_i}, q_{\alpha_i}^*$), where $q_{\alpha_i}$ and $q_{\alpha_i}^*$ correspond to the coordinates of the two particles which can be positioned on the opposite edges of the cell. Under this condition, due to the relation~(\ref{conf}), the integral equations~(\ref{system1}) reduce to algebraic equations in which $\alpha_i$ and $\alpha_j$ are the dummy variables and the indices $i$ and $j$ are fixed
\begin{align} 
		\label{system3a}  
		S_{j,\alpha_i}=\displaystyle\sum\limits_{\alpha_j=0}^2c_{\alpha_j}S_{i,\alpha_j}R_{\alpha_i,\alpha_j},\\
		\label{system3b}
		 S_{i,\alpha_j}=\displaystyle\sum\limits_{\alpha_i=0}^2c_{\alpha_i}S_{j,\alpha_i}R_{\alpha_i,\alpha_j},   
\end{align}   
where 
\begin{align}
	\label{R}
	R_{\alpha_i,\alpha_j}=\exp\left[-\beta\sigma(\alpha_j,\alpha_j)\right].
\end{align}
Unlike our previous work~\cite{Bokun2019}, in which only one equation was considered (due to all $S_{j,\alpha}$ being equal), here in the inhomogeneous case we have a system of equations with a given density distribution across the entire system. 

Given the fact that the screened potential is symmetric with respect to the densities $\alpha_i$ and $\alpha_j$ and depends only on the densities at sites $i$ and $j$, equations~(\ref{system3a})--(\ref{system3b}) form a closed system of equations with respect to~$S_{j,\alpha_i}$ and~$S_{i,\alpha_j}$. As a result, we obtain a closed system of six equations.

Notably, in the case when $\alpha=0,1$, this system is simplified and takes on the form
\begin{align}
	\label{hey44}
	S_{j,0_i}&=\frac{c_{0_i}}{S_{i,0_j}}+\frac{c_{1_i}}{S_{i,1_j}},\\
	\label{hey45}
	S_{j,1_i}&=\frac{c_{0_j}}{S_{i,0_j}}+\frac{c_{1_i}}{S_{i,1_j}}R_{i,j}.
\end{align}   
Equations~(\ref{hey44}) and~(\ref{hey45}) can be closed by the similar equations with indices $i$ and $j$ swapped. In the equation~(\ref{hey45}) we account for the fact that due to the absence of interaction between vacancies, the relation~(\ref{R}) is not equal to one only when $\alpha_i=1$ and $\alpha_j=1$. We, therefore, simplify the notation:
\begin{align}
	{R_{\alpha_i,\alpha_j}}=R_{1_i,1_j}.
\end{align}
For further examination of equations~(\ref{hey44}) and~(\ref{hey45}), let us introduce the notations
\begin{align}
	\label{M}
	M_{ij}&=S_i(0_j)S_j(0_i),\\
	\label{hey48}
	h_{ij}&=\frac{S_{i,1_j}}{S_{i,0_j}},\qquad h_{ji}=\frac{S_{j,1_i}}{S_{j,0_i}}.
\end{align}
Due to definitions~(\ref{M}) and~(\ref{hey48}), the system~(\ref{hey44}) and~(\ref{hey45}) takes on the form
\begin{align}
	\label{hey_M}
	M_{ij}&=c_{0_j}+\frac{c_{1_j}}{h_{ij}},\qquad M_{ji}=c_{0_i}+\frac{c_{1_i}}{h_{ji}},\\
	\label{hey_Mh}
	M_{ij}h_{ji}&=c_{0_j}+\frac{R_{ij}c_{1_j}}{h_{ij}},\qquad M_{ji}h_{ij}=c_{0_i}+\frac{R_{ij}c_{1_i}}{h_{ji}}.  
\end{align}
The solution of equations~(\ref{hey_M})--(\ref{hey_Mh}) can be written as
\begin{align}
	\label{hey411}
	h_{ij}^2+h_{ij}\frac{c_{1_j}-c_{0_i}-R_{ij}(c_{1_i}-c_{1_j})}{c_{0_j}}-\frac{R_{ij}c_{1_j}}{c_{0_j}}=0.
\end{align}
From equation~(\ref{M}) we also have that 
\begin{align}
	M_{ij}=M_{ji}.
\end{align}
Due to equation~(\ref{singlet_f}), we can derive the pair distribution function~(\ref{pair_f})
\begin{align}
	\label{fex}
	F_2^0(1_i,1_j)=c_{1_i}c_{1_j}\frac{R_{ij}}{S_j(1_i)S_i(1_j)}.
\end{align}
Given the notation~(\ref{M}), the expression~(\ref{fex}) can be presented as
\begin{align}
	\label{hey414}
	F_2^0(i,j)=\frac{c_{1_i}c_{1_j}}{M_{ij}}\frac{R_{ij}}{h_{ij}h_{ji}}.
\end{align}
From equation~(\ref{hey414}) one can see that in the lattice approximation the pair distribution function does not depend on single-particle mean potentials $\phi_{i,j}(q_{\alpha_i})$. Therefore, in this approximation the systems~(\ref{system1}) and~(\ref{K}) become partially decoupled allowing to find the solution of equation~(\ref{system1}) followed by equation~(\ref{K}). In the lattice approximation, the latter can be rewritten as
\begin{align}
	\label{hey415}
	S_{j,\alpha_i}^*=\sum\limits_{\alpha_j=0}^1 c_{\alpha_j}{S_{i,\alpha_j}^{*-1}}L_{\alpha_i,\alpha_j},
\end{align}
where 
\begin{align}
	S_{j,\alpha_i}^*&=\varepsilon_{j,\alpha_i}S_{j,\alpha_i},\qquad \varepsilon_{j,\alpha_i}=\exp\left[-\beta\chi_j(q_{\alpha_i}^*)\right],\\
	L_{\alpha_i,\alpha_j}&=R_{\alpha_i,\alpha_j}W_{\alpha_i,\alpha_j},\qquad W_{\alpha_i,\alpha_j}=\exp\left[-\beta\Phi_{sh}(q_{\alpha_i}^*,q_{\alpha_j}^*)\right].
\end{align}
Since the value of $L_{\alpha_i,\alpha_j}$ is different from one, when $\alpha_i=1$ and $\alpha_j=1$, we put $R_{ij}^*=L_{1_i,1_j}$ allowing the system~(\ref{hey415}) to take on the form~(\ref{hey44}) and present the solution in the form~(\ref{hey411}).

\section{Free energy of the inhomogeneous system}

In the inhomogeneous case, the free energy can be expressed as the sum of local contributions. Each term in this series is a functional calculated in accordance with the algorithm outlined for a given density field. In our previous model~\cite{Bokun2019}, the free energy was a function of one variable, namely the mean density in the system. The equilibrium distribution of density across the system is found from the condition of an extremum of the functional. We note that in this case the number of fluctuating parameters is equal to the number of flat monomolecular layers which is proportional to the cubic root of the number of particles in the system. This contribution is important in the critical region, which we do not consider in this work.

The notations~(\ref{hey48}) and~(\ref{hey414}) allow us to write the short-range part of expression~(\ref{hey39}) in the form
\begin{align}
	\label{h51}
	f_i^{sh}&=c_{1_i}\ln Q_{1_i}+c_{0_i}\ln Q_{0_i}=c_{1_i}\ln\frac{Q_{1_i}}{Q_{0_i}}+\ln Q_{0_i}\nonumber\\
	&=c_{1_i}\ln\left[\prod_{j(i)}\frac{S^*_{j,1_i}}{S_{j,1_i}} \frac{S_{j,0_i}}{S_{j,0_i}^*}\right]+\ln\omega_i \left[\prod_{j(i)}\frac{S^*_{j,0_i}}{S_{j,0_i}}\right]\nonumber\\
	&= c_{1_i}\ln\prod_{j(i)}\left(\frac{h_{ji}^*}{h_{ji}}\right)+\frac{1}{2}\ln\omega_i\prod_{j(i)}\left(\frac{M_{ji}^*}{M_{ji}}\right)-\beta c_{1_i}U^{\rm{ext}}(q^*_{1_i}).
\end{align}
For the long-range part of~(\ref{hey39}), considering that $\Phi_{L}\ne 0$ when $\alpha_i\alpha_j=1$, we have
\begin{align}
	\label{h52}
	&V_i^{\beta} =\frac{1}{2}\sum\limits_{j(i)}c_{1_i}c_{1_j}\left[\xi_{ij}+\beta\Phi_{L}(r_{ij})\right],\\
	\label{h53}
	&\xi_{ij} =\int_{0}^{\beta}\rd\beta\Phi_{L}(r_{ij})h_2(r_{ij}),\\
	\label{h54}
	&h_2(r_{ij}) =F_2^0(r_{ij})-c_{1_i}c_{1_j},	
\end{align}
where $h_2(r_{ij})$ is the correlation function and $r_{ij}$ is the distance between sites $i$ and $j$.

When the long-range interaction corresponds to Coulomb interaction, we have
\begin{align}
	\label{h55}
	\beta\Phi_{L}(r_{ij})=\frac{\beta}{\beta_b}\frac{r_{Bb}}{r_{ij}},
\end{align}    
where~$kT_b=1/\beta_b$ is the reference temperature measured in units of the Lennard-Jones well depth, $r_{Bb}=(e^2\beta_b)(4\piup\varepsilon\varepsilon_0 H)$ is the Bjerrum radius measured in units of cell parameter $H$. The distances in equation~(\ref{h55}) and herein are, therefore, measured in units of $H$.

According to~\cite{Bokun2019}, from equations~(\ref{hey411}) and~(\ref{hey414}) we can write 
\begin{align}
	\label{h56}
	R_{ij}&=\exp\left[-\frac{\beta}{\beta_b}\frac{r_{Bb}}{r_{ij}}\exp\left(-\sqrt{\frac{\beta}{\beta_b}}\nu r_{ij}\right)\right],\\
	\nu&=2\sqrt{\piup c(1-c)r_{Bb}}.
\end{align}
From equations~(\ref{h55}) and~(\ref{h56}) we see that it is convenient to measure the parameter $\beta$ in units of $\beta_b$, i.e., $\beta^*=\beta/\beta_b$. As a result, due to equation~(\ref{h55}), the quantity $\xi_{ij}$ in equation~(\ref{h53}) takes on the form
\begin{align}
	\label{h58}
	\xi_{ij}=\frac{r_{Bb}}{r_{ij}}h_{\beta}(r_{ij}),\qquad h_{\beta}(r_{ij})=\int_{0}^{\beta^*}h_2(r_{ij})\rd\beta^*. 
\end{align} 
We should note that in the exponent for $R_{ij}$ in equation~(\ref{h56}) we have a typical Debye--H\"{u}ckel (DH) form for the screening potential of point ions~\cite{Debye1923}. However, there is a considerable difference for the dependence of the inverse Debye radius $\nu$ on the concentration of mobile ions. In the DH theory, this radius is proportional to the square root of concentration $c$ while in the case considered here it is proportional to the square root of the product $c(1-c)$~\cite{Bokun18}. The results of the two theories coincide only when ion concentrations are low. Another important result in the framework of the theory developed was obtained in~\cite{Bokun18} from the condition of equilibrium between bulk and surface properties of mobile ions. As a result, it was shown that the distribution of ion concentration can be presented in the form
\begin{align}
	\label{bokundicaprio2018}
	c_i=\frac{1}{1+[(1-c)/c]\{\exp[-\beta(\delta u_i-e\phi_i-\delta w_i)]\}},
\end{align}  
which looks like a Fermi--Dirac distribution corrected for the interparticle
interaction contribution and concentration dependent multiplier in front of the exponent.
In equation~(\ref{bokundicaprio2018}), $\phi_i$ is the electric potential at lattice $i$, the subscript $i$ indicates the position of the corresponding lattice site, $e$ is the elementary electric charge, $\beta=1/kT$, $T$ is the temperature, $k$ is the Boltzmann constant, $\delta u_i$ is the surface excess part of Coulomb interaction and $\delta w_i$ is the short range part of the interpaticle interaction contribution.

As we will show in the next two chapters, both of the discussed effects are very important for the description of electrophysical properties of a double layer between a solid electrolyte and a hard charged electrode.

\section{Charge and electric field distributions in the region between two plate electrodes}
    
We consider an electrolyte confined between two parallel electrodes and divide the interelectrode region into layers parallel to the plates. As a result, the variables $c_{1_i}$ and $c_{1_j}$ have the meanings of the average concentrations of particles in layers $i$ and $j$, respectively. The total number of layers is $L$, i.e., $i = 1,2,...,L$. The long-range component of the free energy, given by the formula~(\ref{h52}) consists of correlated and non-correlated parts
\begin{align}
	V_i^{\beta}=V_i^c+V_i^{nc}.
\end{align}  
The correlated part is expressed in terms of the correlation function while the non-correlated part corresponds to the energy of Coulomb interaction between capacitor plates modelling molecular layers. It is convenient to calculate this energy via the Poisson equation~\cite{Bokun18}. As a result, we have
\begin{align}
	\label{h62}
	V_i^{nc}&=\frac{1}{2}n\beta^* x_i\sum\limits_{k=2}^i[x_k(i-k)],\\
	\label{h63}
	n&=4\piup r_{Bb},
\end{align}  
where $\beta^*=\beta/\beta_b$, $x_k=c_k-c$, and $c$ is the average concentration of cations in the homogeneous electrolyte. Equation~(\ref{h62}) holds for the case when electrolyte layers are numbered consecutively from $k=2$ to $k=L-1$. The first and the last layers model the electrode plates, whose charges are used as boundary conditions. Hence, the energy of the external electric field at location $i$ is given by the following expression
\begin{align}
	U_i^{\rm{ext}}=n\beta^*x_1x_i(l_s-i).
\end{align}
The plates have opposite charges ($x_1+x_L=0$), $l_s$ is the index of the middle layer. The model, therefore, is analogous to a parallel-plate capacitor. 

As a result, the electric potential induced by the inner electrolyte charges and the electrodes is described by the expressions
\begin{align}
	\Psi_i&=-n\beta^*\sum_{k=2}^{i}[x_k(i-k)],\\
	\Psi_i^{\rm{ext}}&=n\beta^*x_1(l_s-i).
\end{align}   
According to equations~(\ref{h52})--(\ref{h54}), the correlated part of the free energy is expressed in terms of the short-range correlation function. As the first approximation for $V_i^c$, we take  into account only the first neighbors. Due to equation~(\ref{h58}), we can write for the cubic lattice   
\begin{align}
	V_i^c=\frac{r_{Bb}}{r_{ij}}\left[h_{\beta}(r_{i,i-1})+4h_{\beta}(r_{i,i})+{h_{\beta}(r_{i,i+1})}\right],
\end{align}
where $r_{i,j}$ is the distance between the nearest neighbors in layers $i$ and $j$.

We consider the case of small deviations of charge carrier concentrations $c_i$ from the average value~$c$. Expansion of the function~(\ref{h54}) on $x_i$ leads to the equations
\begin{align}
	\label{h68}
	V_i^c=\beta\tilde{J}(x_ix_{i-1}+4x_i^2+x_ix_{i+1}).
\end{align} 
The linear terms in expansion~(\ref{h68}) vanish since in the expression for the total free energy such terms cancel due to the condition of electroneutrality of the system
\begin{align}
	\sum\limits_{i=1}^{L-1}x_i=0.
\end{align} 
An expression similar to~(\ref{h68}) also appears in the expansion~(\ref{h51}). Since in the approximation chosen, the Van der Waals and the correlated Coulomb parts have the same structure, they can be combined by introducing an expansion coefficient $\tilde{J}$.

\begin{figure}[htb]
	\begin{center}
		\includegraphics [height=0.35\textwidth]  {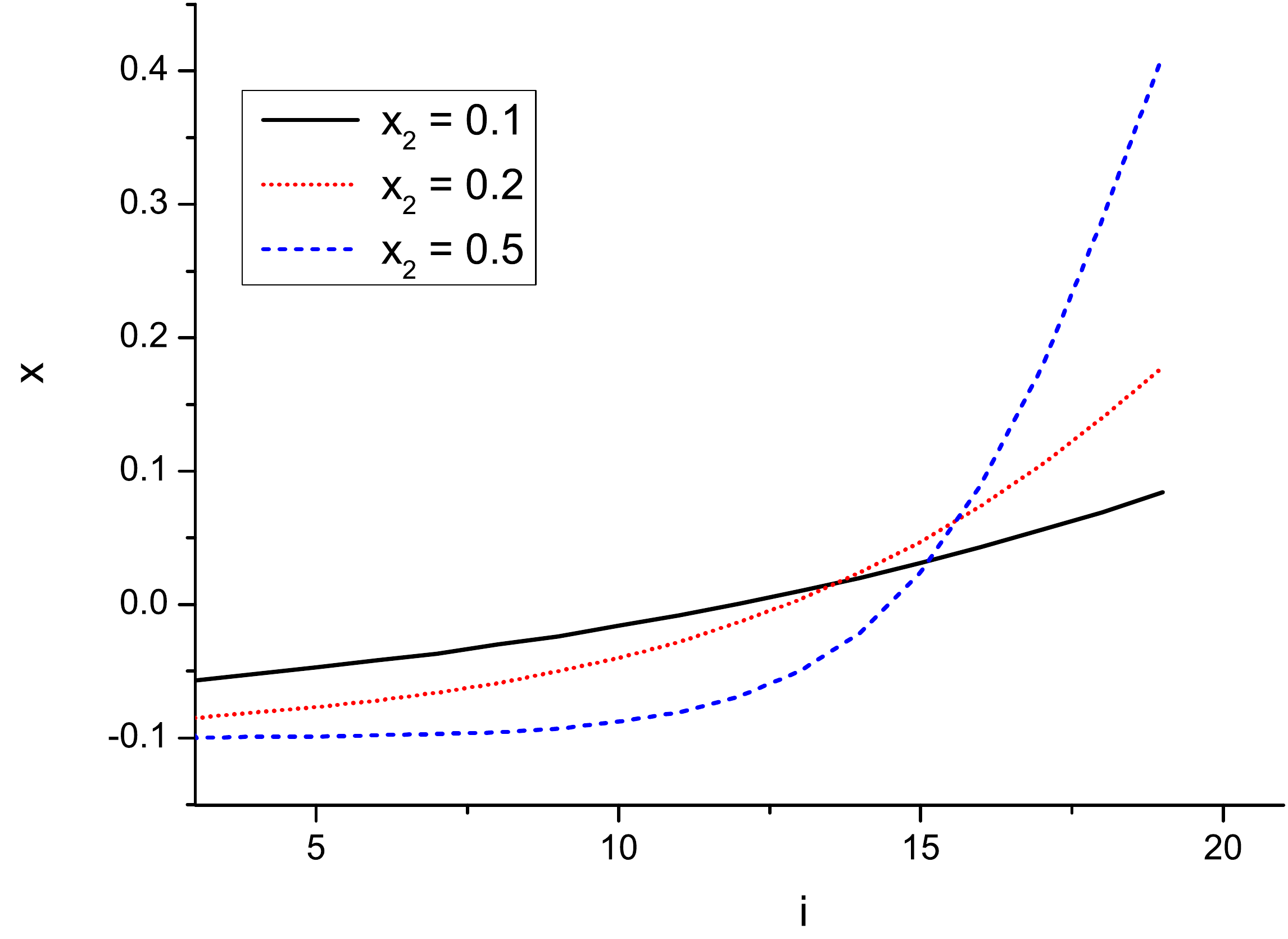}
		\includegraphics [height=0.35\textwidth]  {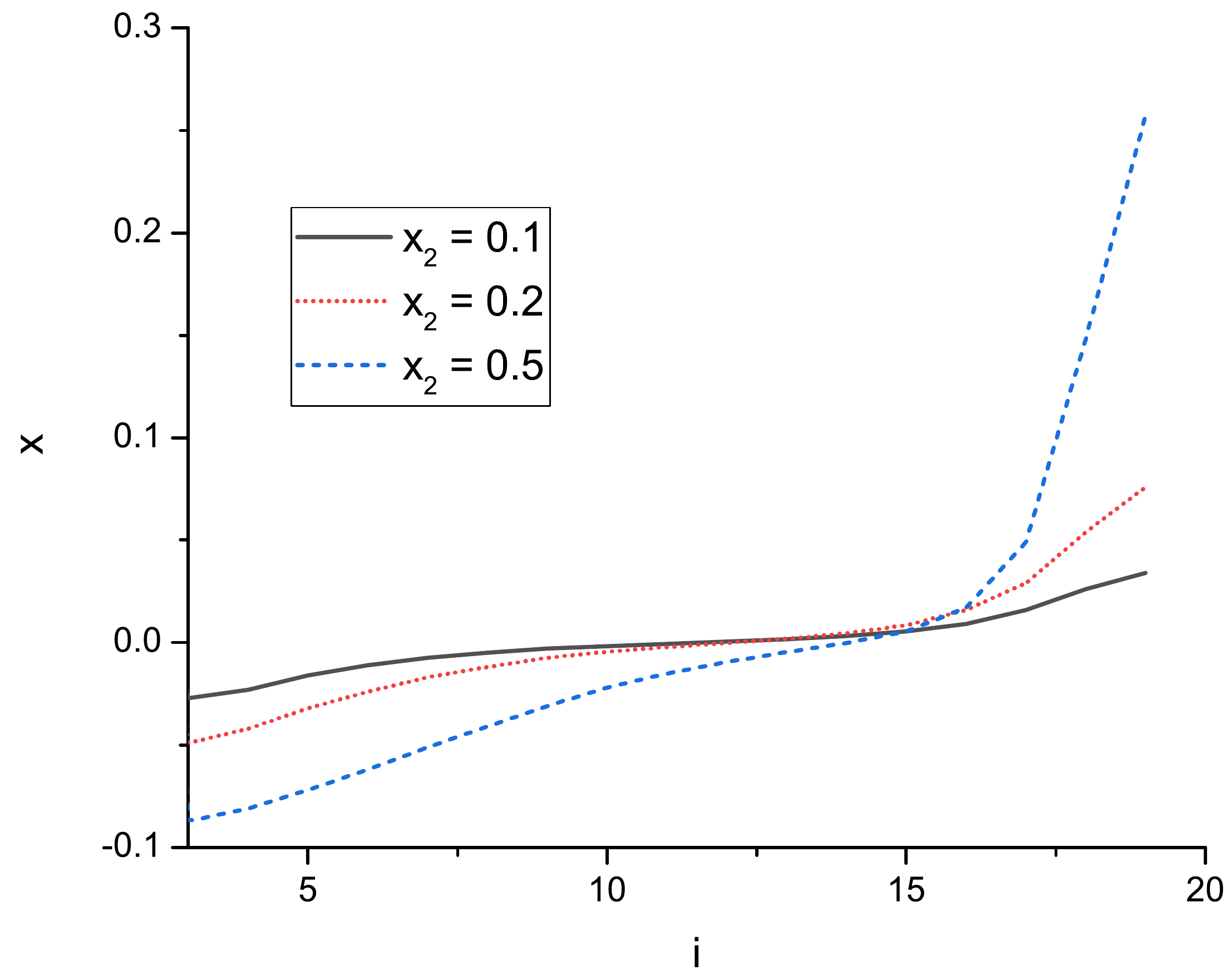}
		\caption{(Colour online) Charge profile for non-interacting ions (left-hand panel) and interacting ions (right-hand panel) in the interelectrode region of 20 layers at different values of electrode plate charge~$x_2$.}
		\label{fig:Fig1}
	\end{center}
\end{figure}

Differentiation of the free energy with respect to the variable $x_i$ leads to an expression for the chemical potential $\tilde{\mu}$ in a local layer $i$. It is convenient to use non-dimensional chemical potential $\mu$ such that $\mu=\beta\tilde{\mu}$ as well as non-dimensional potentials $\Psi = e\beta\tilde{\Psi}$ and $J = \beta\tilde{J}$. This potential consists of a combinatorial part
\begin{align}
	\label{6.10}
	\mu_i^{id}=\ln\left(\frac{c+x_i}{1-c-x_i}\right),
\end{align}   
a part corresponding to non-correlated Coulomb interaction between charges (i.e., an internal electric potential)        
\begin{align}
	\mu_{nc}=n\beta^*\sum_{k=2}^{i}x_k(i-k),
\end{align}
the potential of an external electric field
\begin{align}
	\Psi_i^{\rm{ext}}=n\beta^*x_i(l_s-i),
\end{align}
as well as the potentials of short-range and correlated Coulomb interactions
\begin{align}
	\label{h613}
	\mu_{c_i}&=J\left[6(c+x_i)+(x_{i-1}-2x_i+x_{i+1})\right],\qquad 2<i<(n-1),\\
	\mu_{c_2}&=J\left[4(c+x_2)+(c+x_3)+c\right],\\
	\mu_{L-1}&=J\left[{4(c+x_{L-1})}+(x_{L-2}+c)+c\right].
\end{align}
We perform calculations for the following values of parameters: $\beta^* = 0.5,\ J=-1,\ n=1,\ c=0.1$. The slabs of three different widths $L={21, 41, 61}$ at several values of electrode charges $x_2={0.1, 0.2, 0.5}$ with $x_{L-2}=-x_2$ are considered. The values of variables $x_i (i=2,...,20)$ are determined from minimization of the free energy via the gradient descent method. 

\begin{figure}[htb]
	\begin{center}
		\includegraphics [height=0.35\textwidth]  {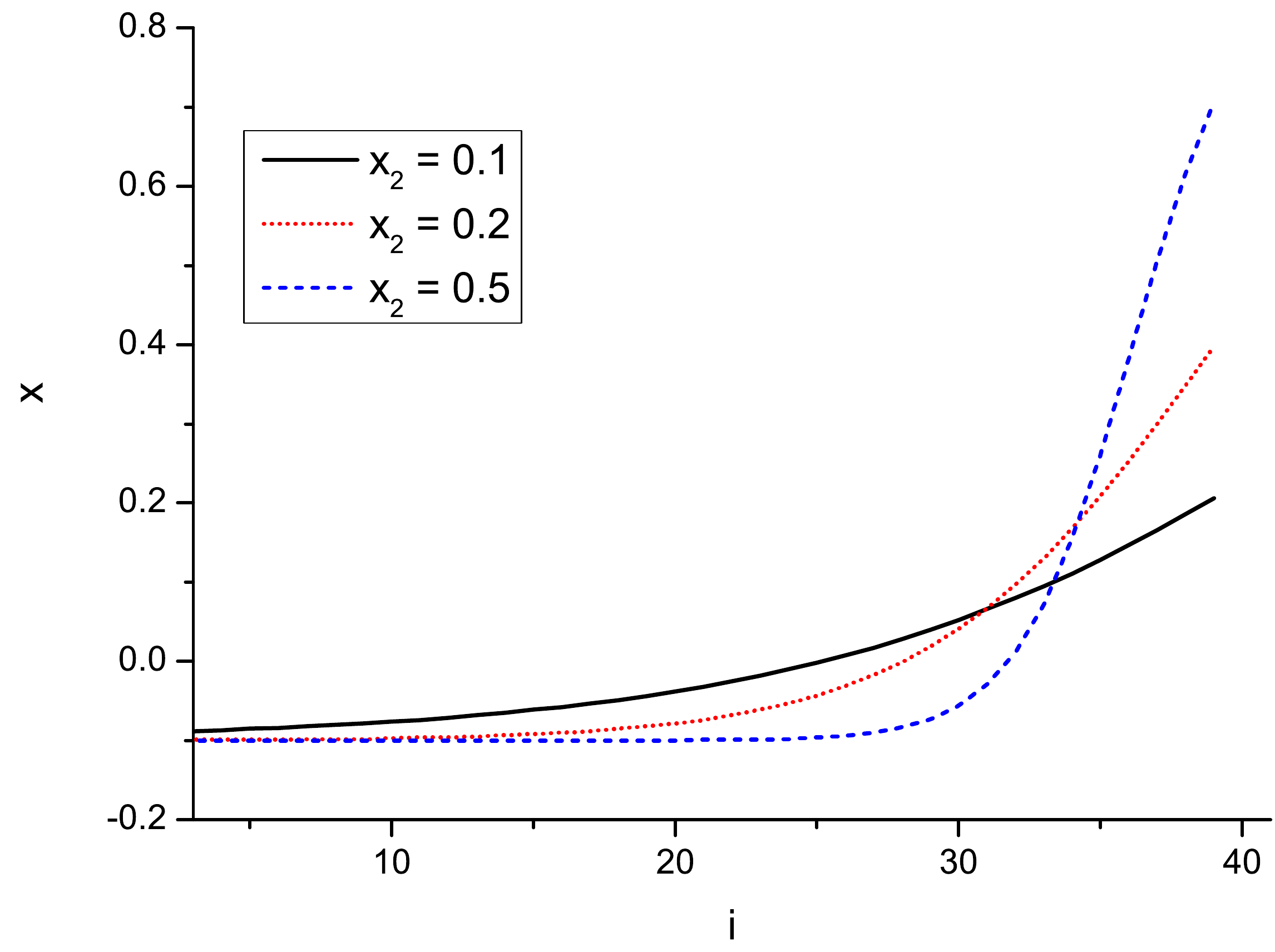}
		\includegraphics [height=0.35\textwidth]  {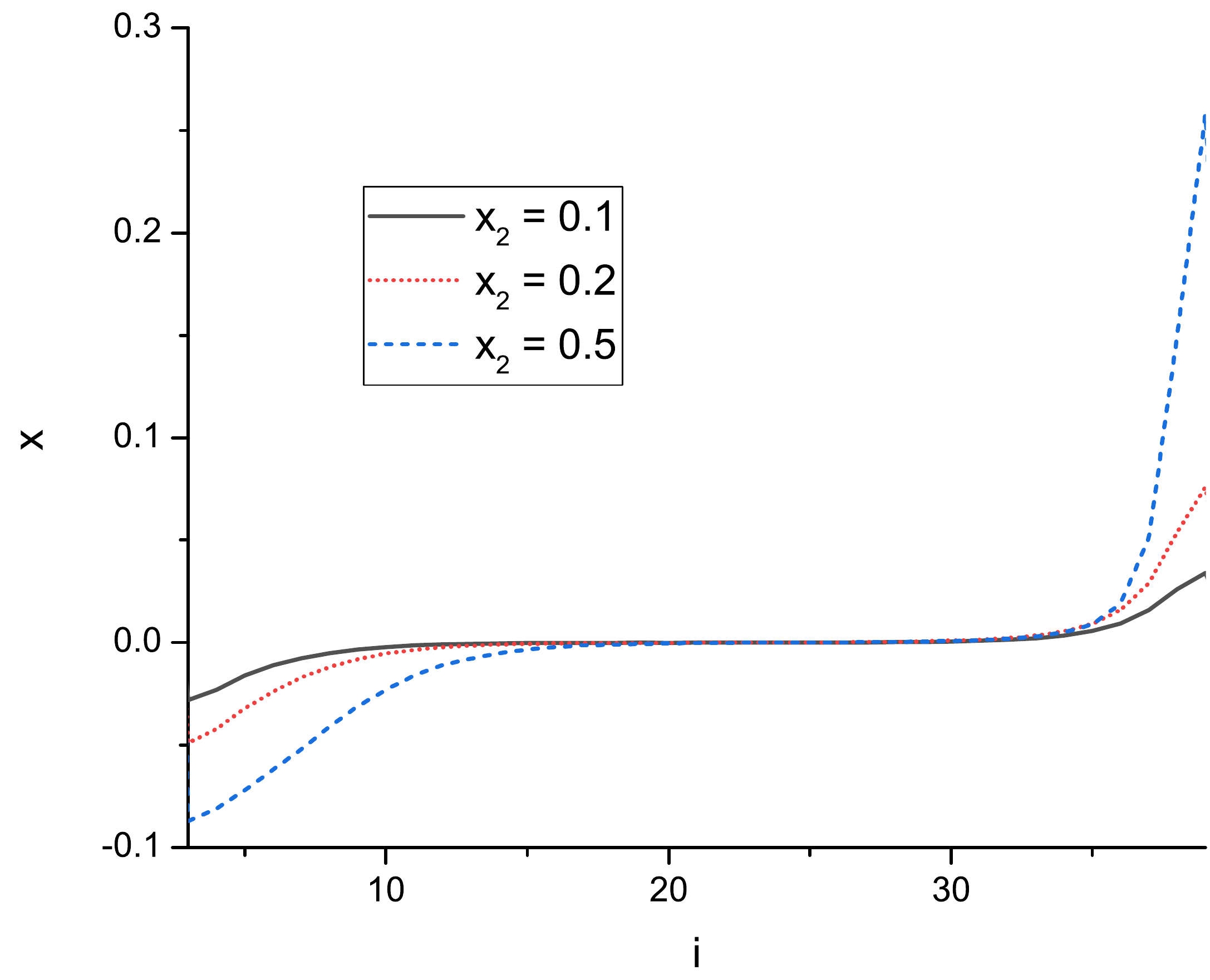}
		\caption{(Colour online) Charge profiles as in figure~\ref{fig:Fig1} for 40 layers.} 
		\label{fig:Fig2}
	\end{center}
\end{figure}

\begin{figure}[htb]
	\begin{center}
		\includegraphics [height=0.35\textwidth]  {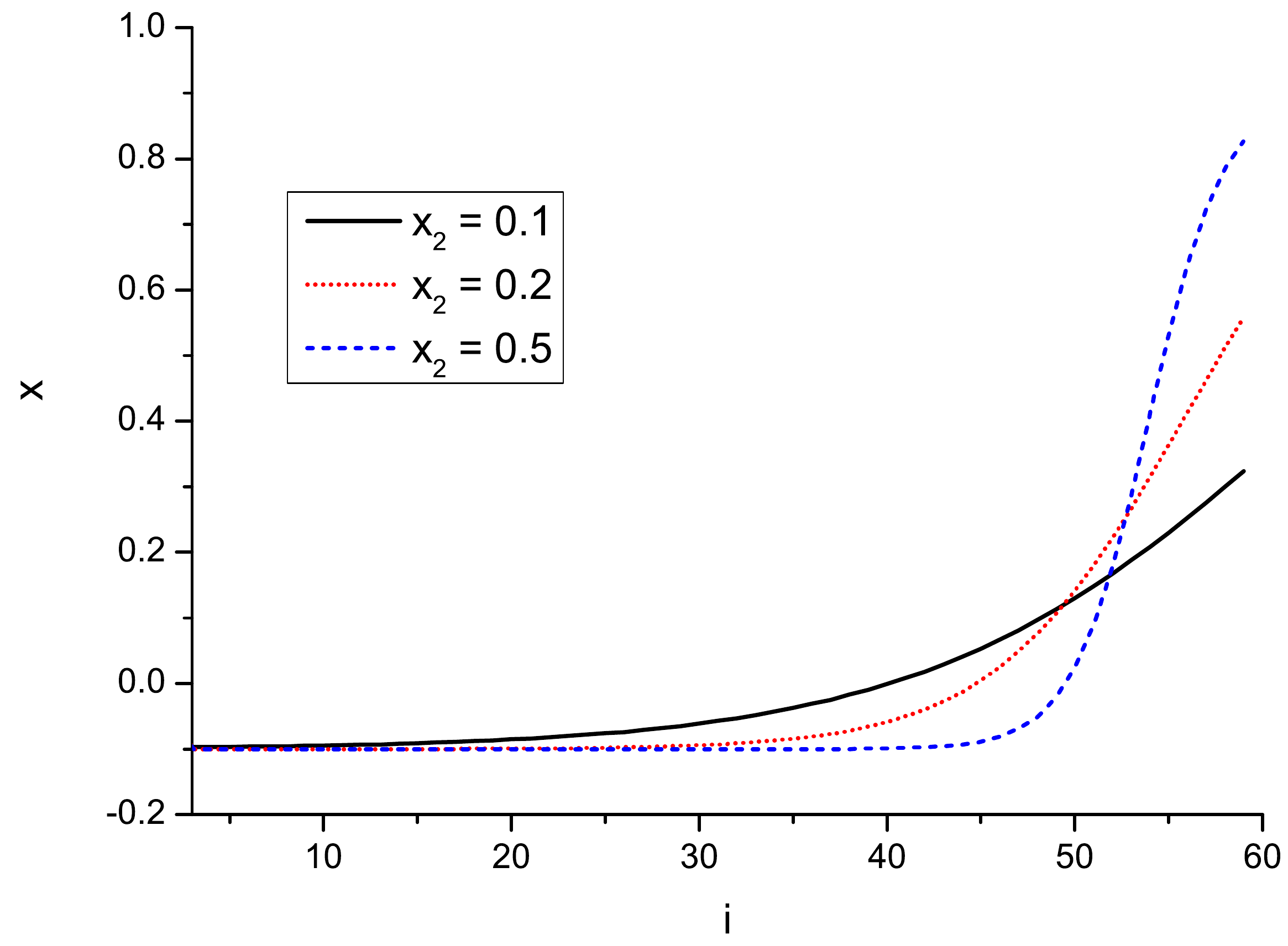}
		\includegraphics [height=0.35\textwidth]  {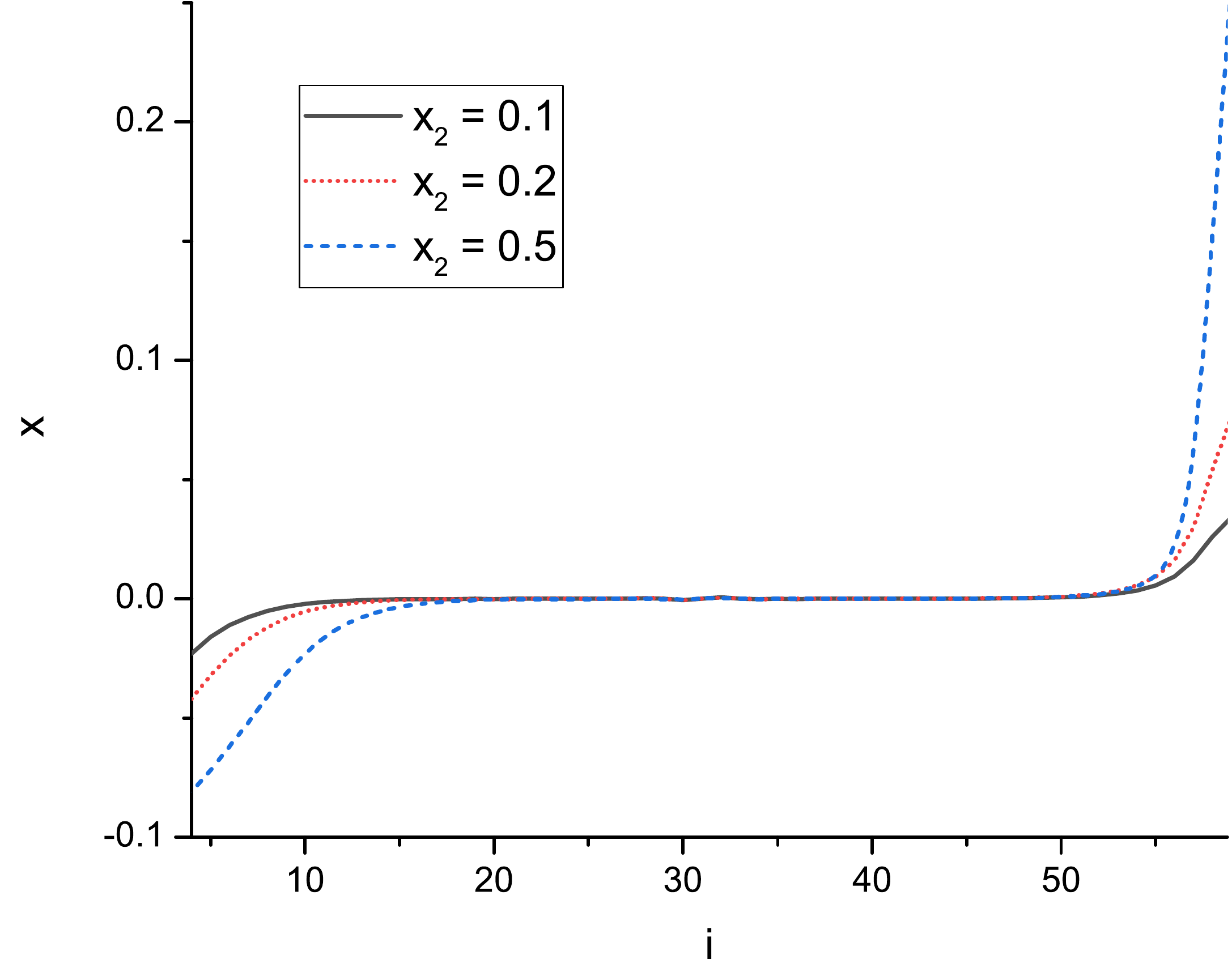}
		\caption{(Colour online) Charge profiles as in figure~\ref{fig:Fig1} for 60 layers.}
		\label{fig:Fig3}
	\end{center}
\end{figure}

Charge distribution in the interelectrode electrolyte film is presented in figures~\ref{fig:Fig1}--\ref{fig:Fig3} in the right-hand panels, while the left-hand panels correspond to the case of non-interacting ions. Due to the concentration dependence of interparticle correlation term $\delta w_i$, the equation~(\ref{bokundicaprio2018}) has a more complex concentration distribution compared with the traditional Fermi--Dirac distribution for non-interacting particles when $\delta w_i=0$. In the considered case, this term is taken into account by the usual iterative procedure. Although the condition of electroneutrality is always satisfied, for the strong charge $x_2=0.5$, the distribution becomes more asymmetric. For the non-interacting particles (figure~\ref{fig:Fig3} on the left), the cations are strongly repelled toward the negatively charged electrode and the non-adjacent electrolyte layers have a nearly constant charge value determined by the average ion concentration $c=0.1$. When interaction is taken into account, the particles are redistributed in a much more symmetric manner. In the latter case, one can also observe the tendency for the formation of a neutral plateau in the middle of the interelectrode region.       

\begin{figure}[htb]
	\begin{center}
				\includegraphics [height=0.35\textwidth]  {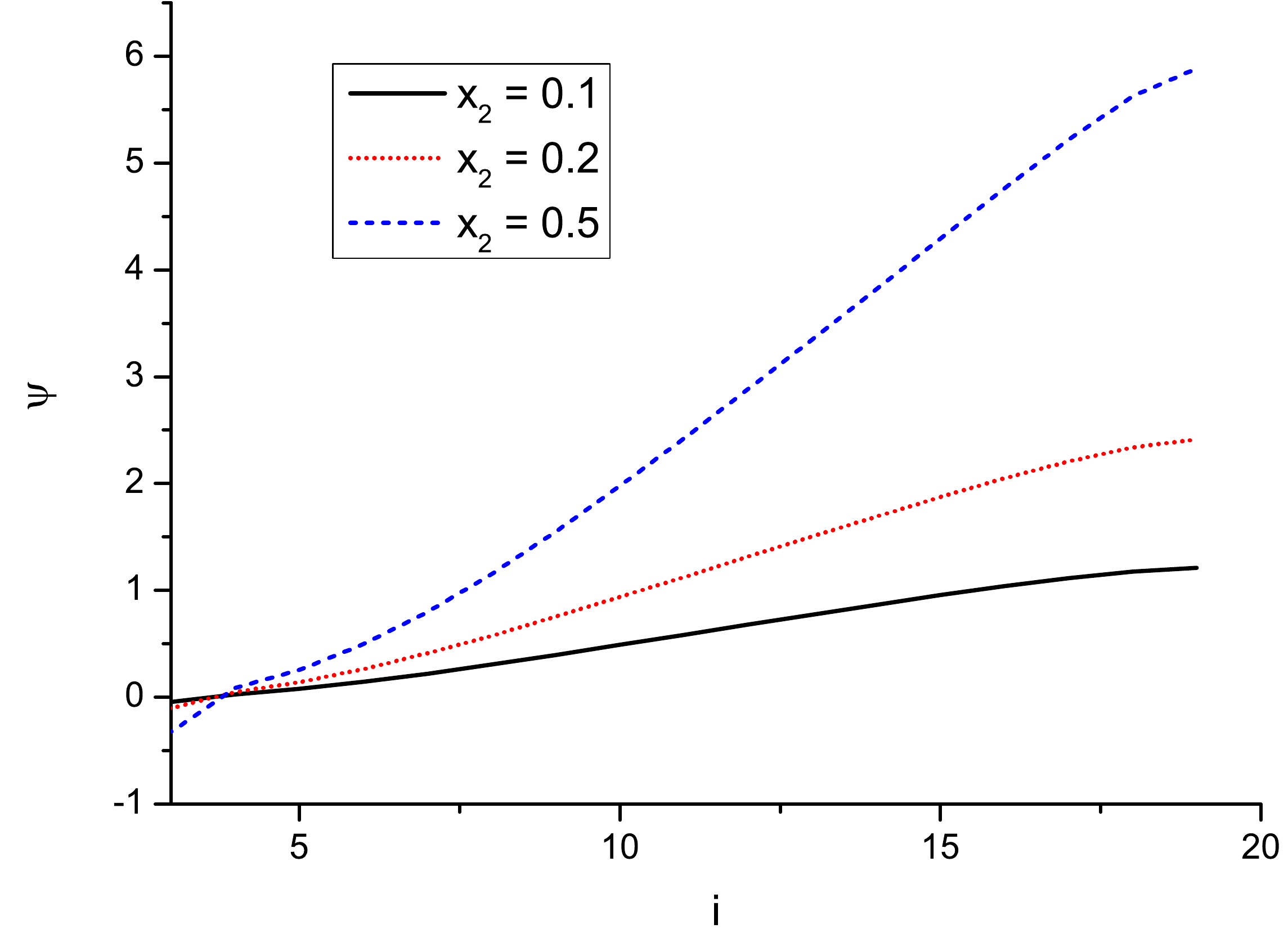}
				\includegraphics [height=0.35\textwidth]  {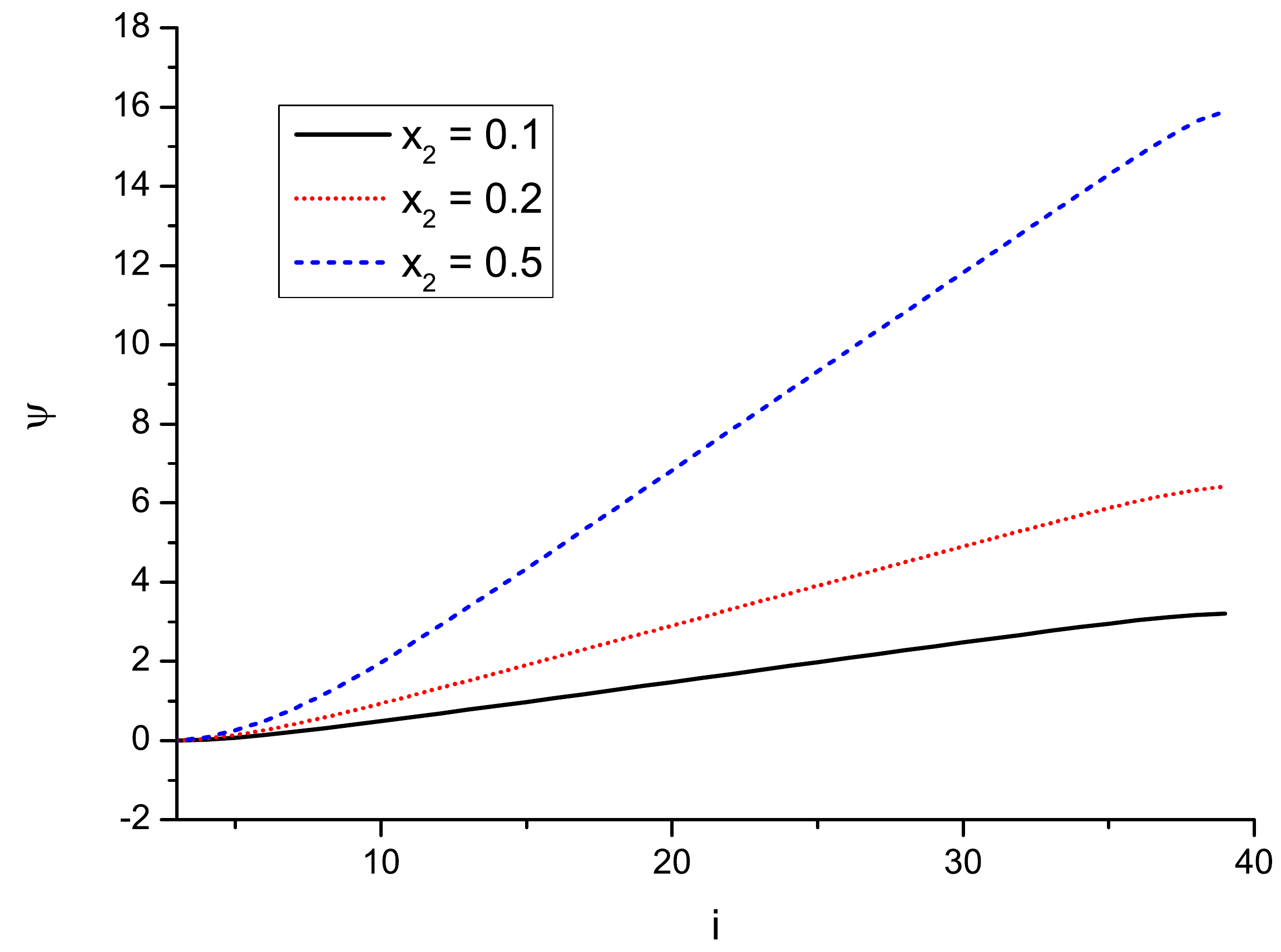}\\
				\includegraphics [height=0.35\textwidth]  {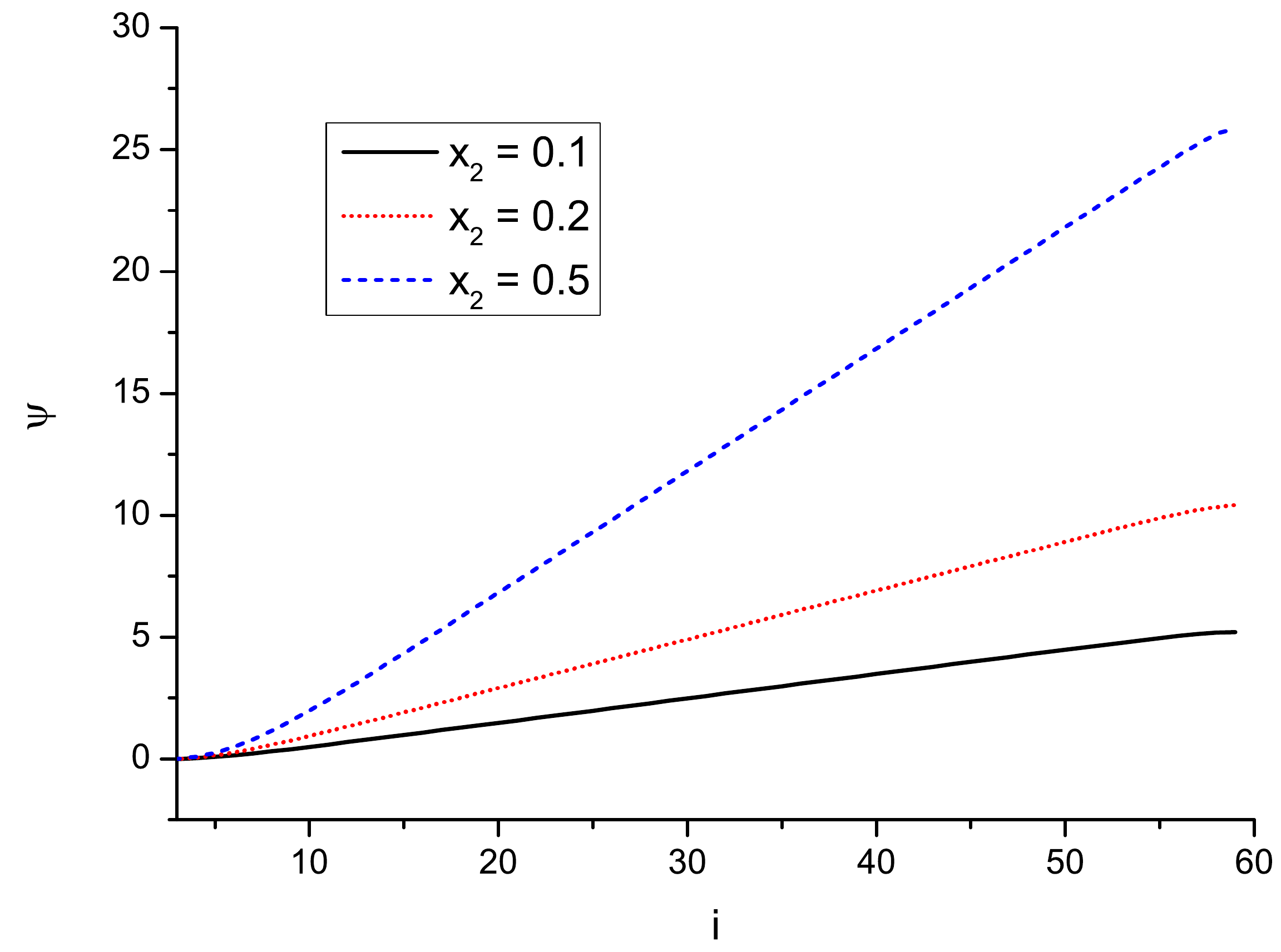}
				\caption{(Colour online) Electric field profile in the interelectrode region of 20, 40, and 60 layers (clockwise starting from top left) at different values of electrode plate charge $x_2$.}
		\label{fig:Fig4}
	\end{center}
\end{figure}

Electric field distributions induced by internal charges in the interelectrode region are shown in figure~\ref{fig:Fig4} for different numbers of layers as well as different plate charges~$x_2$. One can see that the potential profile becomes more linear for wider slabs as well as higher values of electrode plate charges. 

\begin{figure}[htb]
	\begin{center}
		\includegraphics [height=0.35\textwidth]  {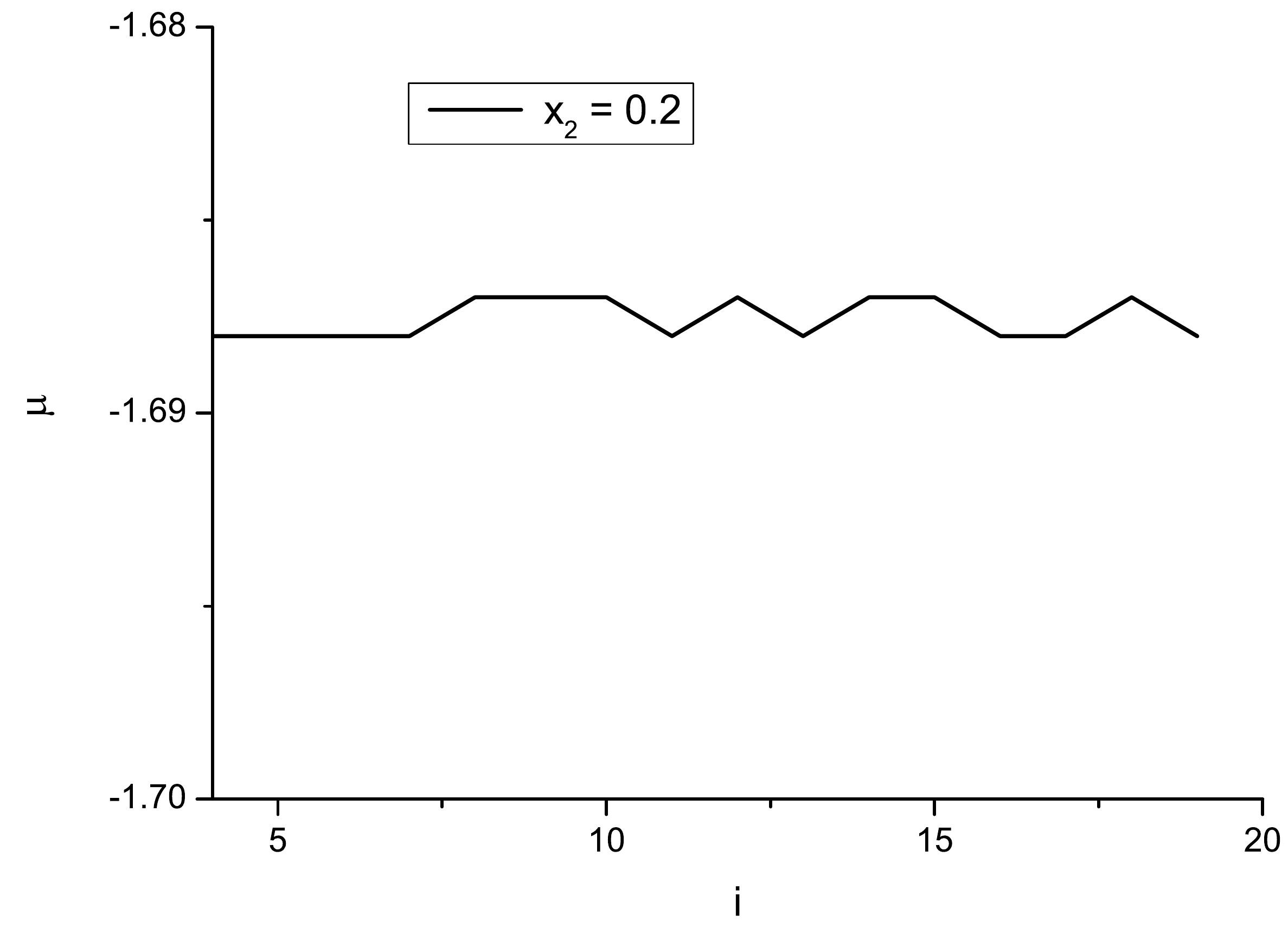}
		\includegraphics [height=0.35\textwidth]  {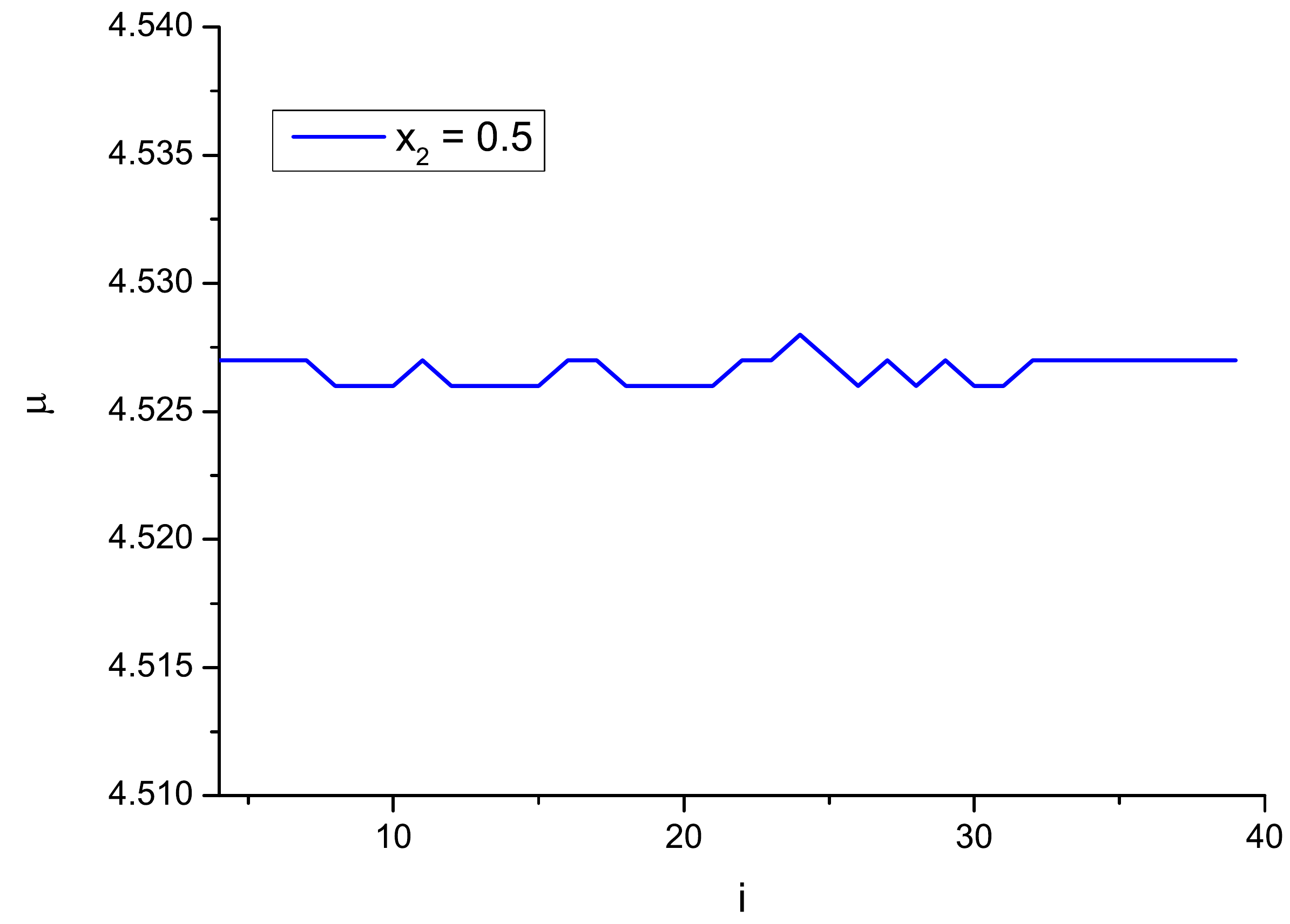}\\
		\includegraphics [height=0.35\textwidth]  {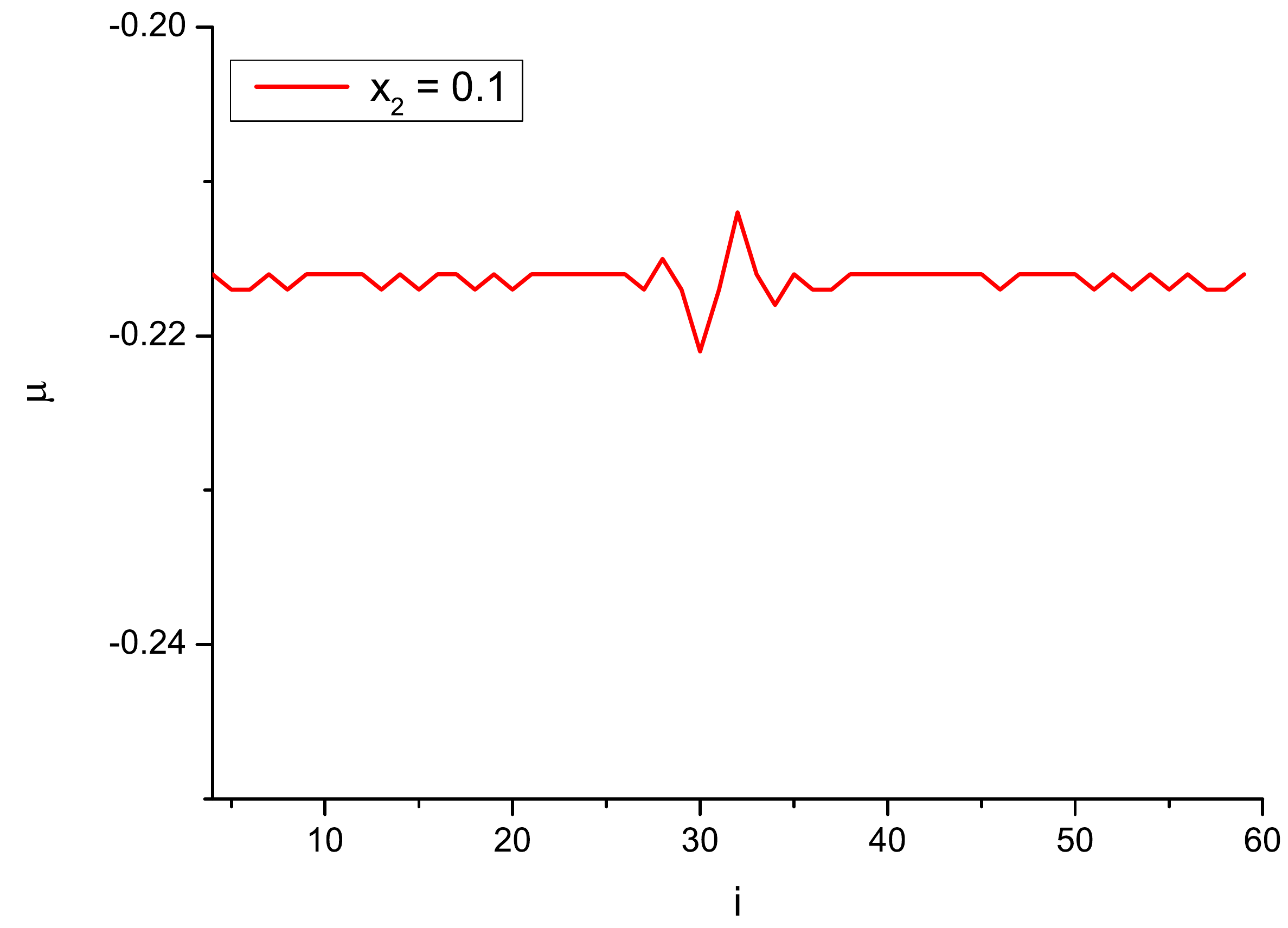}
		\caption{(Colour online) Constant value of the chemical potential in the interelectrode region of 20, 40, and 60 layers (clockwise starting from top left) at different values of electrode plate charge $x_2$.}
		\label{fig:Fig5}
	\end{center}
\end{figure}

The chemical potentials for different values of $i$ are presented in figure~\ref{fig:Fig5}. In each case, we show only the curve for one particular value of $x_2$, noting that a similar quasi-constant behavior takes place for all other values of $x_2$. The results of computations show that the model considered leads to reasonable deviations of the chemical potential from the given constant equilibrium values. For the parameters chosen, the system finds itself in a stable equilibrium state as indicated by the positive value of the determinant of the matrix~$D$
\begin{align}
	D_{ij}=\frac{\partial\mu_i}{\partial x_j}.
\end{align}

\section{Electric capacitance}

The differential electric capacitance is defined as
\begin{align}
	\label{6.17}
	c_{D}=\frac{\delta Q}{\delta \Psi},
\end{align}
where $\delta Q$ is the variation of the charge due to a variation of the potential. 

\begin{figure}[htb]
	\begin{center}
		\includegraphics [height=0.35\textwidth]  {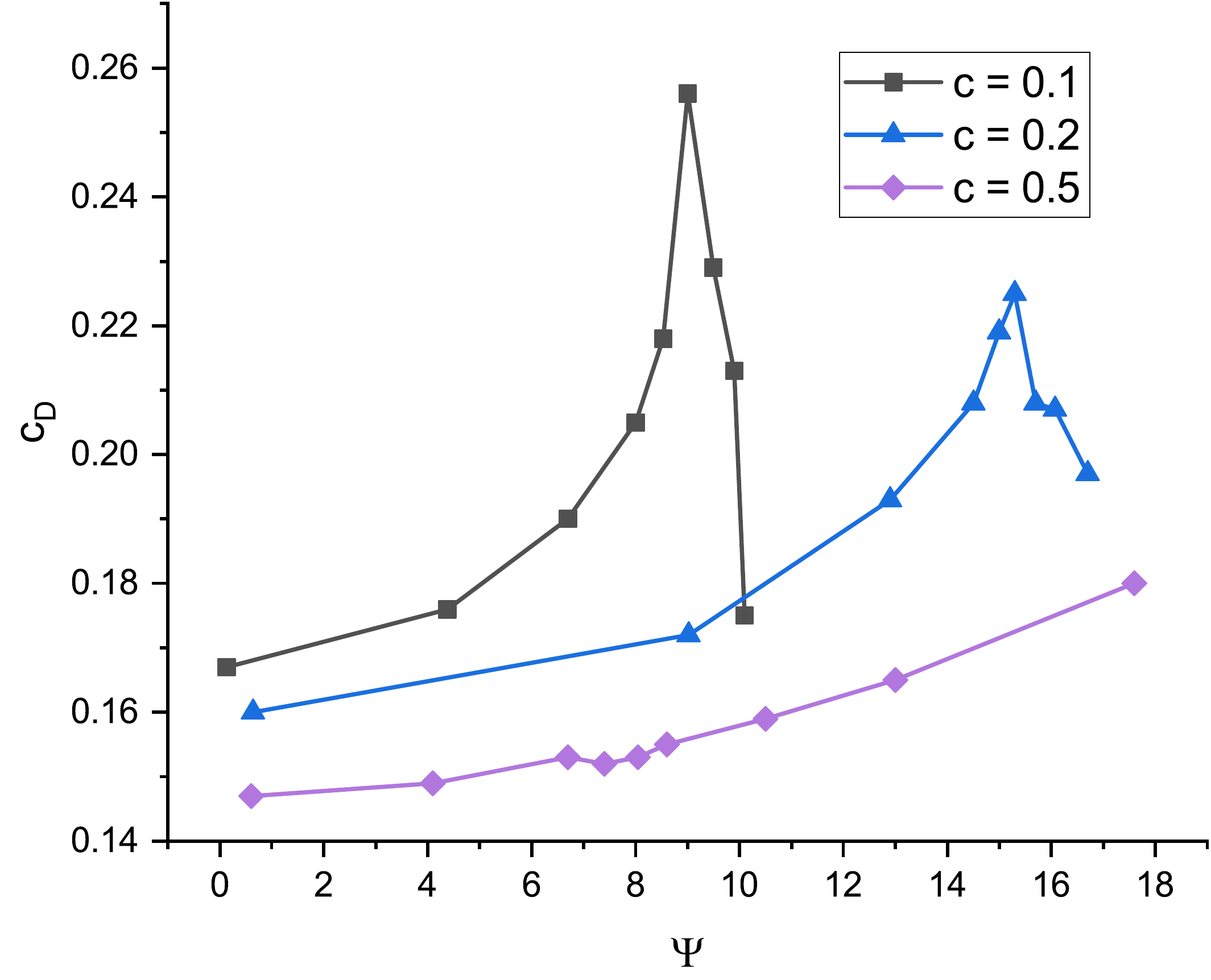}
		\caption{(Colour online) Differential capacitance as a function of the electric potential.}
		\label{fig:Fig6}
	\end{center}
\end{figure}

We can find the charge in the bulk region by summing up the charges in all the layers
\begin{align}
	\label{6.18}
	Q = \frac{1}{2}\sum_{i=1}^{L-1}x_i\textrm{sgn}(x_i).
\end{align}
The potential $\Psi$ is, therefore, the difference of the potentials of the boundary layers
\begin{align}
	\label{6.19}
	\Psi=\beta^*\sum\limits_{k=2}^{L-1}k^*x_k.
\end{align}
To calculate the quantity~(\ref{6.17}), we find the variations~(\ref{6.18}) and~(\ref{6.19}) caused by the variation of the charge of the electrode layers $\delta x_2=h$. The quantities $\delta Q$ and $\delta\Psi$ can be determined as
\begin{align}
	\label{6.20}
	\delta Q = Q(x_2+h)-Q(x_2),\\
	\delta\Psi=\Psi(x_2+h)-\Psi(x_2).
\end{align}
Equations~(\ref{6.18}) and~(\ref{6.19}) tell us that in the case of many layers the quantities~(\ref{6.20}) are the differences of large numbers, which requires high precision when determining these differences from the condition of a constant chemical potential (see figure~\ref{fig:Fig6}). 
For this reason, the accuracy of calculations sharply diminishes at high potentials and at large numbers of variables. We demonstrate the potential dependence of the differential capacitance for the case when the total error of the calculations is approximately five percent (figure~\ref{fig:Fig6}). To get the values of $c_D$ at large $L$ and $\Psi$, we need to address the problem of a decreasing accuracy with increasing values of these parameters. In figure~\ref{fig:Fig6}, we present the diffetential capacitance as a function of the potential at different mean concentrations of the cations. At small values of the potential, the differential capacitance differs very little from the constant value corresponding to the linear theory. Smoother changes of the differential capacitance as a function of the potential with increasing mean concentrations is the consequence of equation~(\ref{6.10}) at $c=0.5$ containing the inflection point, which widens the applicability region for the linear description. This also explains the shift of maxima to the right as the mean concentration increases. A fall of the differential capacitance at high potentials is the result of saturation in the distribution of charges in the regions close to the electrodes. In order to increase the density of cations in one region and decrease in another, one should, therefore, increase the potential as the density of charges in the sample increases. When the potential exceeds some critical value, the charge can no longer go up and the differential capacitance falls to zero as in figure~\ref{fig:Fig6}. It is also worth noting that the curves obtained can be extended to the negative values of the electric potential. For the case under consideration, these curves will be  mirror symmetric with respect to those presented for the positive values of the electric potential. Such a shape of the differential capacitance as a function of the electric potential is known in the literature as camel-like shape and has recently been widely discussed in connection with the room temperature ionic liquids~\cite{Goodwin2017,Jitvisate2018}.

\section{Conclusions}

The paper provides a statistical mechanical description of an inhomogeneous equilibrium state of a solid electrolyte by extending the approach constructed earlier for a homogenous state~\cite{Bokun2019}. As in our previous work, we consider the case of mobile cations in a neutralizing field of fixed anions. The Van~der~Waals interaction is accounted for by means of mean cell potentials while the long-range Coulomb interaction is described in the framework of the collective variables formalism. In addition, the possibility of taking into account particle fluctuations not only around lattice sites but also between the sites is shown. An expression for the free energy as a functional of the particle density field is derived. Similar to~\cite{Patsahan19}, the electric potential distribution and the charge density distribution in the interelectrode region of the electrolyte are computed though with additional, more accurate considerations. Notably, the expression for the chemical potential used in~\cite{Patsahan19}, which is characteristic of ideal solutions, is replaced by the calculation of the free energy functional developed in~\cite{Bokun2019} and extended here for inhomogeneous systems. This enabled us to take into account correlations and interparticle interactions more accurately using mean potentials. Furthermore, in contrast to~\cite{Patsahan19}, the equilibrium states are calculated based on numerical minimization of the free energy and on the enforcement of the condition that the total chemical potential across the system should be constant. Such an approach allowed us to avoid the problems that arose in~\cite{Patsahan19} from the direct solution of the system of equations, which proved to be poorly defined. The results obtained are in qualitative agreement with those reported in~\cite{Ciach01,Archer2007,Zhuang2016,Pini2017}.

An important aspect of the present paper is the use of the results of our previous work~\cite{Bokun18}, namely the expression for the screened potential. This potential is of the same form as that in the Debye--H\"{u}ckel theory but with a different screening radius. In contrast to the Debye--H\"{u}ckel theory, in which the inverse screening radius is proportional to the square root of the concentration of mobile ions $c$, in the case considered it is proportional to the square root of the product $c(1-c)$~\cite{Bokun18}. The results of the two theories coincide only when the concentrations of mobile ions are small. Another principal result of this work was also obtained in~\cite{Bokun18} and comes down to the equation~(\ref{bokundicaprio2018}) for ionic concentration distribution. This expression is presented in the form of the Fermi--Dirac distribution corrected for the contribution from interparticle interaction and a concentration-dependent factor in front of the respective exponent. Both effects proved to be highly instrumental in the description of electrophysical properties of the electric double layer for the model under consideration. Notably, we showed that the electric differential capacitance as a function of the electric potential calculated employing these effects has a typical camel-like shape. At small values of the electric potential, the capacitance gradually increases, reaches a maximum, and then falls sharply as the potential is further increased, which is consistent with saturation of the electric charge in the near-electrode region in accordance with the distribution~(\ref{bokundicaprio2018}). A more gradual change of the capacitance at higher concentrations of mobile ions is due to the fact that the relation~(\ref{6.10}) at $c=0.5$ contains an inflection point that widens the region of linear description. This is also the reason for the shift of the curve maxima and the decrease of their values in the direction of higher potentials.

\section*{Acknowledgement}

This project has received funding from the European Union's Horizon 2020 research and innovation programme under the Marie Sklodowska-Curie grant agreement No. 734276.


\ukrainianpart

\title{Розподіли заряду та потенціалу електричного поля у неоднорідному твердому електроліті, який знаходиться між двома електродами}
\author{І. Кравців\refaddr{label1}, Г. Бокун\refaddr{label2}, М. Головко\refaddr{label1}, Н. Прокопчук\refaddr{label2}, Д. ді Капріо\refaddr{label3}}
\addresses{
	\addr{label1} Інститут фізики конденсованих систем НАН України, вул. Свєнціцького, 1, 79011, Львів, Україна 
	\addr{label2} Білоруський державний технологічний університет, вул. Свердлова, 13а, 22006, Мінськ, Білорусь 
	\addr{label3} Інститут хімічних досліджень Парижу Національного центру наукових досліджень, Париж, Франція
}

\makeukrtitle

\begin{abstract}
	\tolerance=3000%
	
	Розглянуто твердий іонний провідник із катіонною провідністю, який знаходиться у міжелектродній області. Оскільки аніони мають великий розмір, вони вважаються нерухомими і утворюють однорідний компенсуючий електричний фон. Модель може бути використано для опису властивостей керамічних провідників. Для статистико-механічного опису таких систем, які характеризуються короткосяжними взаємодіями типу Ван дер Ваальса і далекосяжними кулонівськими взаємодіями, застосовано підхід, який поєднує метод колективних змінних і метод середніх потенціалів. Цей формалізм було застосовано у попередній роботі авторів [Bokun G., Kravtsiv I., Holovko M., Vikhrenko V., Di Caprio D., Condens. Matter Phys., 2019, \textbf{29}, 3351] для опису однорідного стану і у даній роботі узагальнено на неоднорідний випадок зумовлений наявністю зовнішнього електричного поля. В результаті, середні коміркові потенціали є функціоналами поля густини і можуть бути описані замкнутою системою інтегральних рівнянь. Цю задачу розв'язано у гратковому наближенні і досліджено розподіли заряду та потенціалу електричного поля у міжелектродній області в залежності від заряду на пластинках електродів, а також розраховано та проаналізовано диференційну електроємність.    
	
	\keywords керамічні провідники, середні потенціали, граткове наближення, метод колективних змінних, парна функція розподілу, хімічний потенціал
	
\end{abstract}

\lastpage
\end{document}